\documentstyle[psfig,12pt,epsf]{article}
\newcommand{\sect}[1]{\setcounter{equation}{0}\section{#1}}

\def\be{\begin{equation}}
\def\ee{\end{equation}}
\def\bea{\begin{eqnarray}}
\def\eea{\end{eqnarray}}
\def\bq{\begin{quote}}
\def\eq{\end{quote}}

\renewcommand{\d}{\delta}
\newcommand{\r}{\rho}
\newcommand{\f}{\psi}

\newcommand{\half}{\frac{1}{2}}
\newcommand {\k}{\kappa}
\newcommand{\dt}{\sqrt{-g}}
\newcommand{\p}{\partial}

\begin{document}
\thispagestyle{empty}

\vspace*{1cm}
\begin{center}
{\Large \bf \Large\bf Quantum effects in Acoustic Black Holes:\\ the Backreaction}\\
\vspace*{2cm}
R. Balbinot\footnote{E-mail: balbinot@bo.infn.it}, S. Fagnocchi
\footnote{E-mail: fagnocchi@bo.infn.it}\\
\vspace*{0.2cm}
{\it Dipartimento di Fisica dell' Universit\`a di Bologna}\\
{\it and INFN sezione di Bologna}\\
{\it Via Irnerio 46, 40126 Bologna, Italy  }\\
\vspace*{0.5cm}
and A. Fabbri\footnote{E-mail: fabbria@bo.infn.it}\\
\vspace*{0.2cm}
{\it Departamento de Fisica Teorica, Facultad de Fisica, }\\
{\it Universidad de Valencia, 46100-Burjassot, Valencia, Spain }\\
\vspace*{3cm}
{ \bf Abstract}
\end{center}
We investigate the backreaction equations for an acoustic black hole formed in a Laval
nozzle under the assumption that the motion of the fluid is one-dimensional. 
The solution in the near-horizon region shows that as phonons are (thermally) radiated 
the sonic horizon shrinks and the temperature decreases. This contrasts with the behaviour
of Schwarzschild black holes, and is similar to what happens in the evaporation of 
(near-extremal) Reissner-Nordstr\"om black holes (i.e., infinite evaporation time). Finally, 
by appropriate boundary conditions the solution is extended in both the asymptotic 
regions of the nozzle.

\bigskip
\noindent PACS: 04.62.+v, 04.70.Dy, 47.40.Ki.
\\Keywords: Acoustic black holes, Hypersonic flow, Hawking radiation, Backreaction. 
\vfill
\setcounter{page}{0}
\setcounter{footnote}{0}
\newpage

\section{Introduction}

In 1974 Hawking \cite{hawking} announced to the physical community his famous result: 
black holes are not black at all, but emit thermal radiation at a temperature 
proportional to the horizon surface gravity. 
Hawking came to this remarkable conclusion by examining, within the framework of quantum 
field theory in curved space, the propagation of a massless scalar field in the geometry 
of a collapsing body forming a black hole. If one assumes spherical symmetry, 
the space-time in the region exterior to the body is described by the Schwarzschild solution 
(Birkhoff theorem). 
For these kinds of black holes the horizon is located at $r=2M$ 
($M$ is the black hole mass) and its surface gravity is $k=1/4M$. This yields an emission 
temperature $T_H=\frac{\hbar k}{2\pi}=\frac{\hbar}{8\pi M}$
(the velocity of light and Boltzman constant are set to one). 
As a consequence black holes, 
the final end-state of gravitational collapse according to classical General Relativity, 
are unstable. 
\\The actual evolution of an evaporating black hole should in priciple be 
described (until quantum gravity effects become relevant) by the semiclassical Einstein equations (backreaction equations)
\be 
\label{Gback}
G_{\mu\nu}(g_{\alpha\beta})=8\pi \langle T_{\mu\nu}(g_{\alpha\beta})\rangle\ ,
\ee
where $G_{\mu\nu}$ is the Einstein tensor and the r.h.s. is the expectation value 
of the quantum stress tensor for the matter fields which drive the evaporation.
\\ Eq. (\ref {Gback}) has to be solved for the evaporating black hole metric 
$g_{\alpha\beta}$. This requires the knowledge of $\langle T_{\mu\nu}(g_{\alpha\beta})\rangle$
 for an arbitrary (let's say spherically symmetric) metric.
Unfortunately, because of the extreme difficulty of the problem, no such expression is 
available and the backreaction is usually modelled by extrapolating Hawkings's result, 
which is strictly valid only for static or stationary black holes. 
Therefore a spherically symmetric black hole of mass $M(t)$ is supposed to emit radiation at a temperature $T=\hbar (8\pi
M(t))^{-1}$. Because of this emission, it looses mass at a rate given by the Stefan-Boltzman law $\frac{dM}{dt}\simeq
-AT^4\simeq -1/M^2$, where $A=16 \pi M^2$ is the horizon area.
As the mass decreases the black hole becomes hotter, as can be seen from the temperature 
formula given above, and eventually disappears (with a final explosion?). Its lifetime is of the order $M_0^3$ where $M_0$ is the initial mass of the black hole. 
\\If the black hole possesses a conserved electric charge $Q$ ($|Q|<M$) the picture is quite different. In this case the space-time is described by the Reissner-Nordstr\"om solution for which the surface gravity reads
\be
\kappa=(r_+ -M)/r_+^2
\ee
where $r_+$ is the radius of the horizon, i.e.  $r_+=M+\sqrt{M^2-Q^2}$.
One sees that, for near-extremal black holes ($M\stackrel{\sim}{>}|Q|$), as the mass of the hole decreases because of the evaporation, the temperature $T=\frac{\hbar}{2\pi}k$ drops costantly.
When $M$ is reduced to $M=|Q|$ the temperature is zero.
This final state (remnant) is reached in infinite time. This corresponds to the so called third law of black hole thermodynamics.

In 1981 a remarkable paper of Unruh \cite{unruh81} appeared showing that a quantum emission similar to the 
one predicted by Hawking for black holes is expected in a seemingly completely different physical context, namely fluids
undergoing hypersonic motion. This far reaching result opened a continuously developing field of research (the so called
black hole analogue models \cite{libro}) in condensed matter physics where, unike gravity, the hope to perform experimental test on these theoretical predictions does not seem so remote.
\\In this paper we will give a first  insight on   the backreaction this  emitted radiation has on the fluid.
The plan of the paper is the following.
\\In section II we report Unruh's analysis of the sound  propagating  in a hypersonic fluid within the action formalism which is then used to write the backreaction equations for the fluid motion.
\\In section III we introduce a lower dimensional model where the backreaction equations can be solved to find the first order in $\hbar$ correction to the classical flow.
\\In section IV we evaluate the quantum stress tensor for the classical background to be inserted in the backreaction equations.
\\In section V we solve the backreaction equations near the sonic horizon.
\\In section VI and VII the near horizon solution is extended in the asymptotic regions 
through the imposition of appropriate matching conditions.
\\Section VIII is devoted to the conclusions and some extrapolations of our results.
\\In Appendix A and B we report tecnical details needed to perform the calculations.


\sect{Hypersonic flow and sonic black holes}

Unruh considered an irrotational, homentropic fluid. In this case the Eulerian equations of motion can be derived from the action
\be
\label {Sclassic}
S=-\int d^4 x \left[ \r \dot{\f}+\half \r (\nabla \f)^2+u(\r)\right]
\ee
where $\r$ is the mass density, $\f$ the velocity potential, i.e. $\overrightarrow{v}=\overrightarrow{\nabla}\f$, and $u$ the internal energy density. The overdot means differentiation with respect to (newtonian) time.

Varying $S$ with respect to $\f$ yields the continuity equation
\be
\label{Eqcont}
\dot {\r}+\overrightarrow{\nabla}\cdot (\r \overrightarrow{v})=0\ ,
\ee
whereas variation with respect to $\r$ gives Bernoulli's equation
\be
\label {Ber}
\dot{\f}+\half\overrightarrow{v}^2+\mu(\r)=0\ ,
\ee
where $\mu(\r)=\frac{du}{d\r}$.

One can now obtain the linearized wave equation for the propagation of sound waves in a background mean flow by replacing
\bea
\f&\rightarrow&\f+\f_1\ ,\\
\r&\rightarrow&\r+\r_1 \ ,\nonumber
\eea
where $\r$ and $\f$ define the mean flow; they are assumed to satisfy the equations of motion and $\f_1$ and $\r_1$ are small amplitudes perturbations.
\\ Expanding $S$ up to quadratic order in these perturbations one obtains
\bea
\label{S}
S&=&S_0-\int d^4 x \left[ \r_1 \dot{\f}_1+\half \frac{c^2}{\r} \r_1^2+\half \r (\overrightarrow{\nabla} \f_1)^2+\r_1\overrightarrow{v} \cdot  \overrightarrow{\nabla} \f_1 \right]\\
&=&S_0+S_2\ .\nonumber
\eea
The speed of sound $c$ is defined as 
\be
\label{defc}
c^2= \r \frac{d\mu}{d\r} \ . \ee

One can derive the equation of motion for $\r_1$
\be
\label{romotion}
\r_1=-\frac{\r}{c^2}\left( \dot{\f}_1+\overrightarrow{v} \cdot  
\overrightarrow{\nabla} \f_1\right) \ .
\ee
Since $\r_1$    occurs    quadratically in (\ref{S}) we may use eq. (\ref{romotion}) 
to eliminate it and obtain an action for the potential $\f_1$ only
\be
\label{s}
S_2=-\int d^4 x \left[ \half \r (\overrightarrow{\nabla} \f_1)^2
-\frac{\r}{2c^2} ( \dot{\f}_1 +\overrightarrow{v} \cdot  \overrightarrow{\nabla} \f_1 )^2 
\right] \ .
\ee
The remarkable thing is that $S_2$ can be written as 
\be
\label{sgr}
S_2=-\half \int d^4 x \dt g^{\mu\nu} \p_\mu \f_1 \p_\nu \f_1 \ ,
\ee
where, following Unruh, we have introduced the acoustic metric
\be \label{acmio}
g_{\mu\nu}=-\frac{\r}{c}\left(
\begin{array}{cc}
c^2-v^2&\overrightarrow{v}^T\\
\overrightarrow{v}&-I
\end{array}
\right)
\ee
and $I$ is the three-dimensional identity matrix.\\ 
In the form of eq. (\ref{sgr}) $S_2$ is completely equivalent to the action for a 
massless scalar field $\f_1$ propagating in a curved space-time whose line element is
\be
ds^2=g_{\mu\nu} dx^\mu dx^\nu \ .
\ee
Varying the action (\ref{sgr}) with respect to $\f_1$ gives the sound wave equation which can be written in a simple and elegant geometrical way
$$
\Box \f_1=0 \ , 
$$ 
where $\Box=\nabla_\mu \nabla^\mu$ is the covariant derivative with respect to the acoustic 
metric $g_{\mu\nu}$.
Variation of $S_2$ of eq. (\ref{sgr}) with respect to the acoustic metric gives the so called 
``pseudo energy momentum tensor'' (PEMT) $T_{\mu\nu}$ 
\be
T_{\mu\nu}=-\frac{2}{\sqrt{-g}}\frac{\delta S_2}{\delta g^{\mu\nu}}\ .\ee
As discussed by Stone \cite{stone} the invariance of $S_2$ under diffeomorphisms implies
the covariant conservation of $T_{\mu\nu}$ 
\be
T^{\mu\ ;\nu}_{\ \nu}=0 \ , 
\ee
which describes in a nice compact form the exchange of energy and momentum between the waves
and the fluid reproducing various known laws of classical fluid motion. Of particular interest
are fluids undergoing hypersonic motion.  
In this case the region of the fluid for which $\overrightarrow{v}^2 > c^2$ is called acoustic black  hole,
its boundary $ \overrightarrow{v}^2=c^2$  defines the sonic horizon. 
From this region sound waves cannot propagate upstream since the fluid velocity is bigger 
than the sound velocity. Sound is dragged by the fluid and cannot excape.
\\Using the same arguments of Hawking, Unruh, quantizing the field $\f_1$,
showed that in the formation of a sonic hole one expects a thermal emission 
of phonons at a temperature
\be
T=\frac{\hbar  k}{2\pi  c}\ ,
\ee
where $k$ is the surface gravity of the sonic horizon \cite{visser}
\be
k=\left. \half \frac{d}{dn}(c^2-|\overrightarrow{v}_\bot |^2)\right|_H \ ,
\ee
$n$ is the normal to the horizon.

A tipical situation where this is supposed to occur is a Laval nozzle 
depicted in Fig.1.
\begin{figure}
\centerline{\psfig{figure=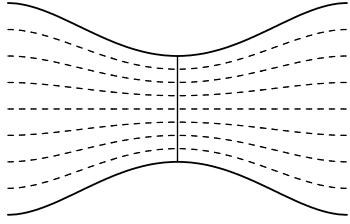,width=2.6in,height=1.7in,angle=-90}}
\caption{A Laval nozzle. The waist of the
nozzle represents the sonic horizon ($|\vec v|=c$). In the region on the right of the waist $|\vec v|<c$
and on the left $|\vec v|>c$ (sonic hole).}
\label{fig1}
\end{figure}  
The fluid flows from right to left; at the waist of the nozzle the fluid velocity reaches the speed of sound. This is the location of the sonic horizon. 
\\Expanding into the left region the velocity increases and the fluid becomes hypersonic; this region describes the acoustic black  hole. In the asymptotic right region a thermal radiation of phonons is expected according to Unruh's analysis.
\\The aim of this paper is to give a first qualitative description of the effects this emitted radiation has on the fluid dynamics, i.e. the backreaction for acoustic black holes.
\\To this end we have to establish the analogue of the semiclassical Einstein equations (\ref{Gback}), i.e. the evolution equation for the fluid driven by the quantum fields.
\\Using standard background field formalism one writes the fundamental quantum 
fields as a sum of background fields ($\rho, \psi$) (not necessarily satisfying the
classical equations of motion) plus quantum fluctuations. Integrating out the quantum
fluctuations one obtains the one-loop effective action formally defined as
\be
\Gamma=S + \frac{1}{2}\hbar Tr\ln\left[\Box_{g(\rho,\vec v)}\right] + O(\hbar^2)\ ,
\ee
where $S$ is the classical action (\ref{Sclassic}) and $\Box_{g(\rho,\vec v)}$ is the
D´Alembertian calculated from an acoustic metric $g_{\mu\nu}(\rho, \vec v)$ of the functional 
form of eq. (\ref{acmio}) \cite{bavili}. We assume the divergences in the determinant of the
above effective action to be removed by employing a covariant regularization scheme. 
This is the key hypothesis of our work. We will comment on it in the conclusion section. 
One can therefore write $\Gamma=S+S_q(g_{\mu\nu}(\rho,\vec v))$, where the quantum part of
$\Gamma$ depends on the dynamical variables $\rho, \vec v$ only through the acoustic metric 
(this is a direct consequence of our hypothesis) and  
coincides with the effective action of a massless scalar field propagating 
in a curved background whose metric is $g_{\mu\nu}(\rho, \vec v)$. 
\\Given this it follows that, using the chain rule,
\bea
\frac{\d S_q}{\d \r}&=&\frac{\d S_q}{\d g^{\mu\nu}}\frac{\d g^{\mu\nu}}{\d \r}
=-\frac{\sqrt{-g}}{2} \langle T_{\mu\nu}\rangle \frac{\d g^{\mu\nu}}{\d \r}
=\frac{\sqrt{-g}}{2} \langle T_{\mu\nu}\rangle \left( \frac{1}{\r}\right)g^{\mu\nu}\\
&=&\half \frac{\sqrt{-g}}{\r} \langle T\rangle =\half \langle T\rangle \frac{\r}{c}\ ,\nonumber
\eea
where 
\be
\langle T_{\mu\nu}\rangle \equiv -\frac{2}{\sqrt{-g}}\frac{\d S_q}{\d g^{\mu\nu}}
\ee
and $\langle T\rangle =\langle T^\mu_\mu\rangle $.\\
Similarly
\bea
\frac{\d S_q}{\d v_i}&=&\frac{\d S_q}{\d g^{\mu\nu}}\frac{\d g^{\mu\nu}}{\d v_i}
=-\frac{\sqrt{-g}}{2} \langle T_{\mu\nu}\rangle \frac{\d g^{\mu\nu}}{\d v_i}=\\
&=&\frac{\r}{c}(\langle T_{ti}\rangle +v_i \langle T_{ii}\rangle ) \ .\nonumber
\eea
In the above expression $g^{\mu\nu}$ is the inverse metric, i.e. 
\be
g^{\mu\nu}=-\frac{1}{\r c}\left(
\begin{array}{cc}
1&\overrightarrow{v}^T\\
\overrightarrow{v}& \overrightarrow{v}\overrightarrow{v}^T-c^2 I
\end{array}
\right)\ .
\ee
\\The backreaction equations are  therefore (taking into account that $g_{\mu\nu}$ does not 
depend on $\dot{\f}$)
\bea
\label{eqbackA}
&\dot {\r}+\overrightarrow{\nabla}\cdot 
(\r \overrightarrow{v})-\nabla_i \left( \frac{\r}{c}(\langle T_{ti}\rangle +v_i 
\langle T_{ii}\rangle ) \right)=0&\ ,\\
\label{eqbackB}
&\dot{\f}+\half\overrightarrow{v}^2+\mu(\r)-  \half \frac{\r}{c}\langle T\rangle =0\ .&
\eea
The $\langle T_{\mu\nu}\rangle $ appearing in the backreaction equations 
should be considered as expectation value of the quantum pseudo 
energy momentum 
tensor operator taken in a quantum state appropriate for  the radiation process, namely the analogue of the Unruh state
\cite{unruh76} (the quantum state suitable to describe black hole evaporation at late times),
in which in the remote past prior to the formation of the sonic hole the quantum field is
in its vacuum state. 
\\Inspection of the backreaction equations reveals that the continuity equation 
(associated to the symmetry $\psi\to \psi + const.$ of the effective action) 
gets modified 
by including the divergence of
the phonons momentum density ($\sqrt{-g}\langle T^0_i\rangle$),
whereas the Bernoulli equation gets an additional contribution to the chemical potential
given by a term
proportional to the trace of the pseudo energy momentum tensor of the phonons.
These equations reflect the underlying two-component structure of the system as in the
Landau-Khalatnikov theory of superfluidity \cite{khalatnikov}.\
\\As in the black hole case the general expression of $S_q$ and 
$\langle T_{\mu\nu}(g_{\alpha\beta})\rangle $ 
is completely unknown and no explicit solution of the backreaction equations (\ref{eqbackA}), 
(\ref{eqbackB}) can be given. 
  
\sect{The dimensional reduction model}

Lower dimensional models are very useful in physics as they allow explicit solutions of the 
dynamical equations to be found, providing at least  a qualitative description of the 
evolution of real four-dimensional systems. 
This attitude has been largely used in studying quantum effects in black hole physics, 
where two-dimensional models enable  investigation of the so called "s-wave sector" 
\cite{strominger}. The most popular one is the one proposed by Callan-Giddings-Harvey-Strominger
(CGHS) \cite{cghs} in which the 4D quantum stress tensor is replaced by a 2D one associated to a
minimally coupled massless scalar field described at the quantum level by the Polyakov action.
This approximation neglects backscattering of the quantum fields caused by the potential barrier
located near the horizon. \\
Within this philosophy a qualitative insight in the backreaction going on in a hypersonic
fluid can be obtained by assuming a one dimensional flow for the fluid; the relevant
physical quantities depend only on the $z$ coordinate running along the axis of the Laval
nozzle of Fig. 1 (i.e. velocity components along $x$ and $y$ are negligible with respect to the
$z$ component). 
\\The effective action for the CGHS-like model for the fluid quantum dynamics can be given as
$\Gamma^{(2)}=S^{(2)}+S_{pol}$, where 
\be 
S^{(2)}=-\int d^2 x A  \left[ \r \dot{\f}+\half \r (\p_z \f)^2+u(\r)\right]  
\ee
is obtained integrating $S$ over the transverse coordinates $x, y$ and 
$A$ is the area of the transverse section of the nozzle. 
Using the chain rule as before one can write 
the backreaction equations as
\bea
\label{back1}
&&A\dot{\r}+\p_z(A\r v)+\p_z \left[ -\frac{1}{c}(\langle T^{(2)}_{tz}\rangle 
+v\langle T^{(2)}_{zz}\rangle )\right]=0
\ , \\
\label{back2}
&&A\left( \dot{\f}+\frac{v^2}{2}+\mu(\r) \right)
-\frac{1}{2} \left(\langle T^{(2)}\rangle  \right)
=0\ ,
\eea
where
\be
\langle T_{ab}^{(2)}\rangle =-\frac{2}{\sqrt{-g^{(2)}}}\frac{\delta S_{pol}}{\delta g^{(2)ab}}
\ee
and $g_{ab}^{(2)}$ is the ($t,z$) section of the acoustic metric (\ref{sgr})
\be
\label{metric2}
g_{ab}^{(2)}=-\frac{\r}{c}\left(
\begin{array}{cc}
c^2-v^2&v\\
v&-1
\end{array}
\right) \ .
\ee 

Our aim is to solve these backreaction equations to first order in $\hbar$ to calculate 
the first quantum corrections to the classical fluid configuration. 
Therefore we need the quantum source terms calculated at zero order in $\hbar$, i.e. for the classical solution.


\sect{The quantum pseudo stress tensor for static sonic black holes}

Before undertaking the study of the backreaction in  eqs. (\ref{back1}), (\ref{back2})  
let us describe the solution of the classical equations of motion which describes a static 
acoustic black  hole and then find the corresponding $\langle T^{(2)}_{ab}\rangle $ 
in the Unruh state 
for this background. The classical equations of motion are given by eqs. (\ref{back1}), 
(\ref{back2}) neglecting the phonon quantum contribution. 
\\Assuming steady flow, i.e. no time dependence, the continuity equation (\ref{back1}) yields
\be
\label{cont2}
A(z)\r(z)v(z)=const \equiv D \ , 
\ee
whereas the Bernoulli equation simplifies to 
\be \label{nbe}
\frac{v^2}{2}+\mu(\rho)=0\ .\ee
Differentiating both equations and taking into account the definition of the velocity of sound
(eq. (\ref{defc})) one easily gets
\be
\frac{A'}{A}=-\frac{v'}{v}(1-\frac{v^2}{c^2})\ ,
\ee
which is the famous nozzle equation. From this one deduces important features of the fluid
motion. First of all the horizon ($|v|=c$) forms at the waist of the nozzle
($A'=0$).\footnote{Note that this is in general not true if external forces are present.
In that case, the equation giving the location of the sonic horizon is $A'/A=-f/\r c^2$,
where $f$ is the external force (see the second of Refs. \cite{garayvisser}). }
Furthermore in the hypersonic region ($|v|>c$) the velocity increases together with the section
of the nozzle, just the opposite of what happens in the more familiare subsonic ($|v|<c$)
case.
\\Let us further assume the sound velocity $c$ constant.
This allows the Bernoulli equation (\ref{nbe}) to be written as
\be
\half v^2+c^2\ln \frac{\r}{\r_0}=0\ ,
\ee
i.e.
\be
\label{rho}
\r=\r_0 e^{-\frac{v^2}{2c^2}}\ .
\ee
Let us choose the profile of the Laval nozzle as 
\be
\label{area}
A=
\left\{
\begin {array}{rl}
A_0+\beta z^2 & \mbox {for}|z|\leq z_0\\
A_+=const & \mbox {for}|z|> z_0
\end{array}
\right.
\ee
where $A_0$, $\beta$ and $z_0$ are constants.
\\From the continuity equation (\ref{cont2}) one has for $|z|\leq z_0$
\be
\label{z2}
z^2=\frac{1}{\beta}\left[ \frac{D}{\r_0}\frac{e^{v^2/2c^2}}{v} -A_0   \right]\ .
\ee
The sonic horizon is located at $z=0$ where $v=-c$ (remember that the fluid is moving 
from right to left). The constant $D$ in eq. (\ref{cont2}) can be evaluated as
\be
-A_0\r_0 e^{-\half}c=D \ ,
\ee
which when inserted in eq. (\ref{z2}) yields
\be
z^2=\frac{A_0}{\beta}\left[ \frac{c}{|v|}e^{\frac{v^2-c^2}{2c^2}} -1   \right]\ .
\ee
This implicitly defines $v=v(z)$ and from eq. (\ref{rho}) we have $\r(z)$; 
so for $|z|\leq z_0$ the acoustic metric $g_{ab}^{(2)}(z)$ of the static sonic black hole 
is determined. \\ \noindent For $|z|>z_0$ the solution is simply $v(z)=\mbox{const}=v(z_0)$
for $z>z_0$ and $v(z)=\mbox{const}=v(-z_0)$ for $z<-z_0$. Similarly for $\r(z)$.
Note that $|v(z)|$ increases as $z$ decreases, while $\r(z)$  decreases as $z$ decreases.
This resulting acoustic metric should be regarded as the asymptotic configuration
of the fluid resulting from the (time dependent) formation of the sonic hole (see for
example \cite{balisovi}).
\\This dynamical evolution of the fluid excites the modes of the quantum field, which is supposed
to be in its vacuum state before the process of formation of the sonic hole begins, 
yielding nontrivial expectation
values for the quantum PEMT. At late times these may be approximated by
$\langle T^{(2)}_{ab}\rangle $ in the Unruh state evaluated 
in this background. \\

We introduce a set of null coordinates
\bea
\label{Unull}
x^-&=&c\left( t-\int \frac{dz}{c+v}\right)\ ,\\
\label{Vnull}
x^+&=&c\left( t+\int \frac{dz}{c-v}\right)\ .
\eea
In terms of these coordinates the acoustic metric is 
\be
g_{ab}^{(2)}=-\frac{\r}{c}\frac{c^2-v^2}{c^2}\left(
\begin{array}{cc}
0&\half\\
\half&0
\end{array}
\right)
\ee 
with $a,b=x_+,x_-$ and the quantum part in eq. (\ref{back1}) reads
\be
\label{Tsum}
-\frac{1}{c}(\langle T_{tz}\rangle +v\langle T_{zz}\rangle )
=-c^2\left[-\frac{\langle T_{--}\rangle }{(c+v)^2}
+\frac{\langle T_{++}\rangle }{(c-v)^2}\right]\ .
\ee
The Polyakov $\langle T_{--}\rangle$ in the Unruh state is \cite{bd}
\be
\label{TUU}
\langle T_{--}\rangle =-\frac{\hbar}{12\pi}C^\half C^{-\half}_{,--}+\frac{\hbar\kappa^2}{48\pi}\ ,
\ee
whereas
\be
\label{TVV}
\langle T_{++}\rangle =-\frac{\hbar}{12\pi}C^\half C^{-\half}_{,++}\ .
\ee
Here the conformal factor of the metric $C$ is
\be
C=\frac{\r}{c}\frac{c^2-v^2}{c^2}
\ee
and $\k$ is related to the surface gravity $k$ on the horizon
\be
k=c^2\k=\left. c\frac{d}{dz}v\right|_H\ .
\ee
For our particular chioce of $A(z)$ (see eq.(\ref{area})), the surface gravity is 
$k=c^2\k=c^2\sqrt{\beta/A_0}$. The trace $\langle T^{(2)}\rangle$ is completely anomalous
and is given by 
\be \label{traza}
\langle T^{(2)}\rangle \frac{\hbar}{24\pi}R^{(2)} \ee
where $R^{(2)}$ is the Ricci scalar for the metric $g_{ab}^{(2)}$. 
\\Note also that in the Unruh state $\langle T_{--}\rangle$ 
is regular on the horizon (where $v=-c$) 
making the r.h.s. of eq. (\ref{Tsum}) finite.
\\Explicit calculation of the components of $\langle T_{ab}^{(2)}\rangle$ 
are given in Appendix I. We recall here only the asymptotic limits
\bea \label{flujo}
\left. \langle T_{--}\rangle \right|_{z\rightarrow \infty}&=&\frac{\hbar\kappa^2}{48\pi}\ ,\\
\left. \langle T_{++}\rangle \right|_{z\rightarrow 0}&=&-\frac{\hbar\kappa^2}{48\pi}\ .\nonumber
\eea


\sect{The backreaction equations near the horizon}

As said before our aim is to find the first order correction in $\hbar$ to the 
classical sonic black hole solution $\r(z),v(z)$ described in eqs. 
(\ref{rho})-(\ref{z2}). To this end we define the quantum corrected velocity potential and density by
\bea
\f_B&=&\f(z)+\epsilon \f_1(z,t)\ ,\\
\r_B&=&\r(z)+\epsilon \r_1(z,t)\ ,
\eea
where the subscript ``B" stays for backreaction, $v=\p_z \f$ and we have introduced a dimensionless 
expansion parameter $\epsilon=\hbar/(|D|A_0)$.  
These values are found by solving the backreaction eqs. (\ref{eqbackA}), 
(\ref{eqbackB}) linearized in $\epsilon$
\bea
\label{EQ1}
&&\epsilon\left[A\dot{\r}_1+\p_z\left(A(\r_1 v+\r\p_z \f_1)\right)\right]=
c^2 \p_z\left[ -\frac{\langle T_{--}\rangle }{(c+v)^2}+
\frac{\langle T_{++}\rangle }{(c-v)^2}\right]\equiv \epsilon F_2(z)\ , \\
 \label{EQ2}
&& \epsilon \left[A(\dot{\f}_1+v\p_z\f_1+\frac{c^2}{\r}\r_1)\right] 
=\frac{\langle T^{(2)}\rangle }{2}
\equiv \epsilon G_2(z)\ . 
\eea
The explicit expressions for $F_2$ and $G_2$ are given in Appendix I.\\
From eq. (\ref{EQ2}) one can obtain $\r_1$
\be
\label{rho1}
\r_1=\frac{\r}{c^2}\left[ \frac{G_2}{A}-\dot{\f}_1-v \f_1' \right]\ , 
\ee
where we indicate $'\equiv \p_z$.\\
Taking the time derivative of this equation and substituting it in the continuity 
eq. (\ref{EQ1}) one obtains the following wave equation for $\f_1$
\bea
\label{EQvisser}
&-\frac{1}{v} \ddot{\f}_1-2\dot{\f}_1'-v'(1+\frac{c^2}{v^2})\f_1'+(\frac{c^2}{v}-v)\f_1''=&\nonumber\\
&=\left[\frac{c^2 F_2}{D}-\left(\frac{G_2}{A}\right)'\right]\equiv H(z)\ .& 
\eea
The l.h.s. is simply proportional to $\nabla_a^{(2)}(\frac{A\rho}{c}\nabla^{(2)a}\f_1)$ 
and the r.h.s. $H(z)$ represents the 
quantum source evaluated for the classical static solution. \\ \noindent
We are mainly interested on the fate of the sonic horizon, so we will solve the wave 
equation (\ref{EQvisser}) near $z=0$. To this end we need an expansion of the 
background quantities $\r$ and $v$, and hence 
$\langle T^{(2)}_{ab}\rangle $, for $\kappa z\ll 1$. 
The results of this tedious calculation are given in Appendix II. 
We give here only the final results. Eq. (\ref{EQvisser}) becomes
\bea
\label{EQvisser1}
&-\frac{1}{c}(1+\kappa z) \ddot{\f}_1-2\dot{\f}_1'-2c\kappa (1+\frac{2}{3}\kappa z)\f_1'-2c\kappa z\f_1''=&\nonumber\\
&=H_0+ H_1 \kappa z\ ,& 
\eea
where
\bea
\label{H}
H_0&\simeq& 0.038 \frac{\kappa^3 c^2 A_0}{\pi}\ , \\
H_1&\simeq& 1.128  \frac{\kappa^3 c^2 A_0}{\pi }\ .
\eea
We shall write the solution in the following form
\be
\label{psi1}
\f_1=f_0(t)+\k z f_1(t) +\frac{\k^2 z^2}{2}f_2(t)+...
\ee
where $f_i(t),i=1,2,3...$ are functions of $t$ only, to be determined.\\
The zero order (in $\k z$) equation is
\be
\label{zeroorder}
\frac{1}{c}\ddot{f}_0(t)-2\k \dot{f}_1(t)-2c\k^2 f_1(t)=H_0\ , 
\ee
whereas the first order equation is 
\be
\label{firstorder}
\frac{1}{c}\left[ \ddot{f}_0(t)+\ddot{f}_1(t)-2\k\dot{f}_2(t)\right]-4c\k^2\left[ f_2(t)+\frac{1}{3}f_1(t)\right]
=H_1\ .
\ee
We shall now define the boundary conditions on the solution. We require that at some given 
time, let's say $t=0$, the evaporation is switched on, i.e. we require 
$\f_1(t=0)=v_1(t=0)=\r_1(t=0)=0$. 
\\The vanishing of $\f_1$ and $v_1=\p_z\f_1 $ at $t=0$ implies from eq. (\ref{psi1}) 
that $f_i(t=0)=0$. Remembering now the relation between $\r_1$ and $\f_1$ 
(see eq. (\ref{rho1})) we have
\bea
\r_1(t=0)=0 &\Rightarrow& \dot{\f}_1(t=0)=\frac{G_2}{A} \qquad\qquad \mbox{i.e.}\nonumber\\
\k^i\frac{\dot{f}_i(t=0)}{i!}&=&\left(\frac{G_2}{A}\right)_i \qquad\qquad i=0,1,2.....
\eea
where the expression on the r.h.s. means the coefficients of the $i$-th term of the 
expansion in $z$ of $G_2/A$.\\
Evaluating eqs. (\ref{zeroorder}), (\ref{firstorder}) and the time derivative of eq. 
(\ref{zeroorder}) at $t=0$ one can obtain the following approximate expression for $\f_1$ 
(see Appendix II for details) 
\be
\f_1=a_1 t+\frac{a_2}{2} t^2+\frac{a_3}{3!} t^3+....+\k z (b_1 t+\frac{b_2}{2} t^2+....)+\frac{\k^2 z^2}{2}(c_1t+...)\ .
\ee
The numerical coefficients are reported in the same Appendix.

Since we have calculated the quantum source only for the static classical metric there is no time evolution of the source
itself. This limits the validity of our solution for small values of $t$ only ($c\k t\ll 1$). For this reason we have just given a power expansion of the functions $f_i(t)$. Therefore we are only able to predict how the backreaction starts, i.e. to connect the static fluid configuration at time $t_0=0$ to a quasi static fluid configuration at time $t_0+\Delta t$. With this in mind we can now give the quantum correction to the velocity field
\be
v_1(z,t)=\k b_1 t+(\k z)c_1\k t=\k t (b_1+c_1\k z)\ .
\ee
Note that $b_1>0$ (see  Appendix II) and being $\k z  \ll 1$ for the validity of our 
solution, we can conclude that $v_1>0$. The background velocity is negative, so we conclude 
that the backreaction has  the net effect to decrease the modulus of the velocity, 
i.e. the fluid is slowing down.\\
The quantum corrected velocity is therefore
\be
v_B=v+\epsilon v_1=-c+c\k z -\frac{1}{6}c\k^2 z^2+\epsilon(b_1+c_1\k z)\k t \ .
\ee
\\The acoustic horizon is defined $|v_B|=c$; this yields 
\be
z_H=-\frac{\epsilon \k b_1 t}{c\k+\hbar c_1\k^2t}\simeq -\frac{\epsilon  b_1 t}{c}\ .
\ee
This equation shows that the horizon is moving to the left with respect the classical location $z_H=0$. Therefore, as the evaporation proceeds, the hypersonic region gets smaller and smaller. 
This is the behaviour one would have naively expected.
The coefficient $b_1$ determining the quantum 
correction to the velocity and hence the evolution of the horizon is just the gradient
of the additional chemical potential related to the expectation value of the trace evaluated at $z=0$. This should be
compared to the black hole case where the evolution of the horizon is determined by the energy flux
($\dot M \propto \langle T^r_{\ t}\rangle$ in spherical symmetry).  
While in the latter case Hawking radiation occurs at the expense of the gravitational energy of 
the black
hole, in the fluid phonons emission takes away kinetic energy from the system.
\\Finally, from eq. (\ref{rho1}) we can calculate $\r_1$
\be
\r_1=\frac{\r_0 e^{-\half}}{c^2}(\alpha+\delta\k z)t\ , 
\ee
where
\bea
\alpha&=&-a_2+c\k b_1\ , \\
\delta&=&-b_2-a_2+c\k c_1 \ .
\eea
This yields $\r_1<0$, i.e. $\r_B=\r+\epsilon\r_1$ is decreasing.


\sect{The backreaction equation in the region $z>z_0$}

In the region $z>z_0$, assuming $\k z_0\ll 1$, the background solutions can be approximated as
\bea
v&=&-c+c\k z_0\ ,\\
\r&=&\r_0 e^{-\half}(1+\k z_0)\ .
\eea
The backreaction equation (\ref{EQvisser}) becomes the simple flat space wave equation
\be
\label{EQflat}
\p_-\p_+\f_1=0
\ee
Since in this region $\langle T^{(2)}\rangle =\langle T_{+-}\rangle =0$ 
and $\langle T_{--}\rangle =const$, then $F_2=G_2=0$.\\
The null coordinates $x^-,x^+$ are defined as follows (see eqs. (\ref{Unull}), (\ref{Vnull})) 
\bea
\label{Uflat}
x^-&=&c\left(t-\frac{z}{c\k z_0}\right)\ ,\\
x^+&=&c\left(t+\frac{z}{2c-c\k z_0}\right)\ .
\eea
The most general solution of (\ref{EQflat}) is a linear combination of two arbitrary 
functions of the kind $\xi(x^-)$ and $\eta(x^+)$. 

The source term given by the quantum pseudo stress tensor is discontinuous in $z=z_0$, 
so our model resembles a sandwich of space-time regions glued across  singular hypersurfaces.
We can find a solution of (\ref{EQflat}) for $z>z_0$ by requiring continuity for the 
$v_1(z,t)$ field across the boundary $z=z_0$. 
Because of the discontinuity in $\langle T_{ab}^{(2)}\rangle $,
there is no way to make $v_1$ and $\r_1$ simultaneously continuous across $z_0$. 
Our choice leads to a quantum corrected acoustic metric wich is continuous in $z=z_0$ up to a conformal factor.\\
For $z<z_0$ we have
\be
v_1(z,t)=(b_1+c_1 \k z)\k t
\ee
and therefore
\be
v_1(z_0,t)=(b_1+c_1 \k  z_0)\k t \ . 
\ee
As seen from the curved space point of view, 
the $z=z_0$ surface is a timelike surface of the acoustic metric of eq. (\ref{metric2}). 
It is crossed in the future only by outgoing rays.\\
Therefore for $z>z_0$ the solution of the wave equation assumes a retarded ``Vaidya form" 
$\f_1=\f_1(x^-)$. Now from eq. (\ref{Uflat})
\be
v_1=\p_z\f_1(x^-)=-\frac{1}{\k z_0}\p_-\f_1\equiv F(x^-)\ . 
\ee
Taking into account that $x^-(t=0,z=z_0)=-1/k\equiv x^-_0$, $F(x^-)$ can be approximated as
\be
F(x^-)=l(x^--x^-_0)\Theta(x^--x^-_0)+O((x^--x^-_0)^2) \ , 
\ee
where $l$ is a constant to be fixed by requiring continuity in $z=z_0$:
\be
\left. F(x^-)\right|_{z=z_0}=lct=v_1(z_0,t)=(b_1+c\k z_0)\k t \ , 
\ee
so that
\be
l=\frac{\k}{c}(b_1+c_1\k z_0)\ , 
\ee
which is positive definite.\\
Therefore for $z>z_0$, $x^->x^-_0$, we have
\be
v_1=v_1(x^-)=\frac{1}{c}(\k b_1+c_1\k^2 z_0)(x^--x^-_0)+...\ , 
\ee
whereas for $x^-<x^-_0$ it vanishes identically.\\The quantum corrected velocity field reads therefore, for $z>z_0$,
\be
v_B=v+\hbar v_1=-c(1-\k z_0)+\frac{\k}{c}(b_1+c_1\k z_0)(x^--x^-_0)\ ,
\ee
showing a velocity decreasing in modulus  with the advanced time $x^-$.\\
Integrating $v_1=\p_z\f_1(x^-)$ one can get $\f_1(x^-)$
\be
\f_1=-\frac{\k z_0}{2c}(\k b_1+c_1\k^2 z_0)(x^--x^-_0)^2
\ee
and from it
\be
\r_1=\frac{\r}{c^2}(-\dot{\f}_1-vv_1)
=\frac{\r_0 e^{-\half}}{c^2}(1+\k z_0)(\k b_1+c_1\k^2 z_0)(x^--x^-_0)\Theta(x^--x^-_0)\ , 
\ee
which increases in terms of the advanced time.
\\As already said $\r_1$ cannot be made continuous across $z=z_0$. 
Evaluating the limits from both sides of the hypersurface $z=z_0$ one can find the jump in $\r_1$
\be
\Delta \r_1=\frac{\r_0 e^{-\half}}{c^2}[a_2+\k z_0(\k b_1+b_2+a_2)ct]
 \ . \ee


\sect{The backreaction equation in $z<-z_0$}

In this region  the backreaction equation is  $\p_-\p_+ \f_1=0$, as in the previous case. Now the null coordinates are
\bea
x^-&=&c\left(t+\frac{z}{c\k z_0}\right)\ ,\\
x^+&=&c\left(t+\frac{z}{2c+c\k z_0}\right)\ .
\eea
As one can see from the acoustic metric (\ref{metric2}), the surface $z=-z_0$ is now a spacelike surface; it is crossed in
the future direction from both ingoing $x^+=const$ and outgoing $x^-=const$ characteristics of the wave equation. So the
solution will depend on both $x^-$ and $x^+$. 

The classical background quantities are now for $z<-z_0$
\bea
\label{vz0}
v&=&-c-c\k z_0\ , \\
\label{rhoz0}
\r&=&\r_0 e^{-\half}(1-\k z_0)\ . 
\eea
The boundary conditions we will impose to find the solution in this region are the 
continuity for both $v_1$ and $\r_1$. Unlike the $z=z_0$ case, here this is possible being $z=-z_0$ a spacelike surface.   Now we impose continuity of $v_1$
\be
\label{psi1z0}
\p_z\f_1(-z_0^-)=\p_z\f_1(-z_0^+)\equiv \varphi_1(ct)\ ,
\ee
where the r.h.s. is calculated on the solution obtained in Section 5. 
Continuity of $\r_1$ reads
\be
\left. \frac{\r}{c^2}[-\dot{\f}_1-v\p_z\f_1]\right|_{z=-z_0^-}=
\frac{\r}{c^2}\left[\frac{G_2}{A}-\dot{\f}_1-v\p_z\f_1\right]\ . 
\ee
Being the background quantities (\ref{vz0}), (\ref{rhoz0}) continuous across $z=-z_0$ 
and taking account of eq. (\ref{psi1z0}) we require
\be
\p_z\f_1(-z_0^-)
=\left. \left(\dot{\f}_1-\frac{G_2}{A}\right)\right|_{z=-z_0^+}\equiv c\varphi_2(ct)\ .
\ee
From Section 5
\bea
\varphi_1(ct)&=&\frac{\k}{c}(b_1-c_1\k z_0)t\ , \\
c\varphi_2(ct)&=&\left( \frac{a_2}{c}-\frac {b_2\k z_0}{c}\right) ct
\eea
and using standard technique for solving the wave equation one eventually arrives at
\bea
v_1&=&\left[ \frac{\k^2z_0}{2c}(b_1-c_1\k z_0)-\frac{1}{2c^2}(a_2-b_2\k
z_0)\right]\left[(x^--x^-_0)\Theta(x^--x^-_0)\right. \nonumber \\
&&\left. -(x^+-x^+_0)\Theta(x^+ -x^+_0)\right]
+\frac{\k}{c}(b_1-c_1\k z_0)(x^- -x^-_0)\Theta(x^--x^-_0)\ , 
\eea
where we have defined
\bea
x^-_0&=&x^-(z=-z_0,t=0)=\frac{1}{\k}\\
x^+_0&=&x^+(z=-z_0,t=0)=-\frac{z_0}{2+\k z_0}\ . 
\eea
So we have a Vaidya ingoing solution for $x^-<x^-_0$
\be
v_1=-\left[ \frac{\k z_0}{2c}(\k b_1-z_0\k^2 c_1)-\frac{1}{2c^2}(a_2-\k z_0
b_2)\right](x^+-x^+_0)\Theta(x^+-x^+_0)\ , 
\ee
whence for $x^->x^-_0$
\bea
v_1&=&\left[ \frac{\k z_0}{c}(\k b_1-c_1\k^2 z_0)-\frac{1}{c^2}(a_2-b_2\k z_0) \frac{z+z_0}{\k z_0(2+\k z_0)}
+ \right.\nonumber\\
&&\left. +\frac{\k}{c}(b_1-c_1\k z_0)(x^--x^-_0)\right]\ . 
\eea
In a similar way we obtain
\bea
&&\r_1 =\frac{\r}{c^2}\left\{ (b_1-c_1\k z_0)(x^--x^-_0)\Theta(x^--x^-_0) 
 +\left[ \frac{\k^2z_0}{2}(b_1-c_1\k z_0)- \right.\right. \nonumber\\ 
 && \left.\left. -\half(a_2-b_2\k
z_0)\right]
\left[(x^--x^-_0)\Theta(x^- -x^-_0)-(x^+ -x^+_0)\Theta(x^+ -x^+_0)\right]\right\}
.\;\;\;\;\;\;\;\;\;\;
\eea


\sect{The fate of the acoustic black hole}

The aim of this paper was to face the backreaction problem for acoustic black holes.\\
First we have found the equations for the fluid driven by the linear quantum fluctuations, 
i.e. the linearized ``backreaction equations". 
\\In a lower dimensional case, keeping the Polyakov approximation for the 
stress tensor, we have solved
these equations near the horizon, under the assumption that the quantum source 
corresponds to the pseudo stress tensor of the quantum fields evaluated for the static classical background. 
\\The information we can gain is just preliminary: we obtain an indication on how the quantum solution starts to depart from the classical configuration, once the quantum effects are switched on. 
\\It's not possible to do anything better even for the gravitational black hole 
evaporation. However very interesting
indications seem to emerge even from this simplified model.  \\
We have shown that the sonic horizon moves to the left with respect to the nozzle and the hypersonic region shrinks.
From the expression of the quantum corrected fluid velocity
\be
v_B=-c+c\k z -\frac{1}{6}c\k^2 z^2 +\epsilon(b_1+c_1\k z)\k t
\ee
and the equation for the horizon
\be
\label{zH}
z_H=-\epsilon\frac{b_1}{c} t
\ee
one can evaluate how the emission temperature of these quasi static configuration varies 
with time, i.e.
\bea
 T&=&\left. \frac{\hbar c}{2\pi }\frac{\p v_B}{\p z}\right|_{z_H}=\nonumber\\
&&=\frac{\hbar c}{2\pi  }\left( \k-\frac{1}{3}\k^2z_H+\frac{\epsilon \k^2 c_1}{c}t\right)\ . 
\eea
This, using eq.(\ref{zH}), yields
\bea
\label{Tsonic}
T&=&\frac{\hbar c}{2\pi }\k \left[ 1+\frac{\epsilon \k}{c}\left(\frac{b_1}{3}+c_1\right)t\right]=\nonumber\\
&&=\frac{\hbar c}{2\pi }\k \left[ 1-\frac{563}{720\pi}\epsilon \k^3 c A_0t\right]\ .
\eea
This expression is rather significative: it tells us that the emission temperature of the sonic black hole decreases in time. Unlike Schwarzschild black holes, the sonic black hole gets cooler as it evaporates. Its behaviour is therefore  much more similar to a Reissner-Nordstr\"om black hole.

It's very interesting to try to extrapolate this behaviour in time. From eq. (\ref{Tsonic}) 
we have
\be
\Delta T\propto-\k^4 c^2 A_0 \Delta t\simeq \propto -T^4\Delta t \ . 
\ee
This yields approximatively
\be
t\sim \frac{1}{T^3}\ . 
\ee
This implies that as the sonic black hole radiates, its temperature decreases reaching 
asymptotically zero value in an infinite time. 

For a near extremal Reissner-Nordstr\"om black hole ($r_+\simeq M\simeq |Q|$) a similar analysis yields
\bea
\frac{d M}{d t}&\simeq & -A_HT_H^4\propto -Q^2(M-Q)^2 Q^2\ , \\ 
\frac{1}{M-Q}&=&t\sim \frac{1}{T^2}\ . 
\eea
So sonic balck holes resemble  near extremal R-N black holes. 
The end-state of the evaporation process corresponds in both cases to a zero temperature  configuration, 
reached asymptotically in time.
\\One should mention that other analog models like a thin film 
of ${}^3 He\mbox{-A}$ with a moving domain wall \cite{jacobson} seem to show 
a non-vanishing end-temperature of the evaporation process.
\\In a similar way one can treat sonic black holes formed by a Bose-Einstein condensate \cite{garayvisser}. 
The major difference comes from the fact that the 
sound velocity $c$ is not constant, but proportional to $\r^\half$. This will be discussed
elsewhere. 


\sect{Conclusion}

The basic question which the analysis presented here leaves unanswered is to what extent
does the behaviour we have obtained for the quantum corrected evolution depend on the
linear dispersion relation used (free scalar field to describe phonons), which ignores short
distance corrections due to the molecular structure of the fluid.
The same uncertainty exists for any study in curved space quantum field theory which involves 
high frequency modes like the Hawking effect or in general renormalization calculations. 
One does not know how the yet unknown behaviour of the spacetime at the Planck scale may
influence the results. So far many studies have been devoted to investigate the robustness of
the thermal radiation predicted by Hawking both in the gravitational context and in the
analogue models. Looking at 2D models one has mainly analyzed the behaviour of the Bogolubov
$\beta_w$ coefficient giving the created particles number once nonlinear
dispersion relations are introduced in the high frequency regime. This is supposed 
to model the breaking of local Lorentz (or Galilean, in the analogue model context) invariance here
expected. The scale at which this happens is the Planck scale for gravity, whereas
for sonic black holes it is the intermolecular size. 
As a result of these studies there is now some consensus (see however the recent paper
\cite{unsc} where interesting counterexamples are given) that the $\beta_w$ coefficient is
basically unaffected by the dispersion relation \cite{jacobson2}. However no one knows yet 
what happens to observables like $\langle T_{\mu\nu}\rangle$, which are in addition sensible
to the regularization scheme used. 
In a pure 2D context and for massless scalar fields, Jacobson \cite{jacobson3} has argued that,
within a covariant regularization, no significant deviation from the usual expression for the
trace anomaly (eq. (\ref{traza})) and the flux (eq. \ref{flujo})) are expected 
if one introduces a cutoff at high frequencies.
However, for the hydrodynamical system we have considered, covariance is a symmetry of the 
phonons low energy effective theory only, which is broken 
at short distance.  
Hence non covariant terms depending on the microscopic physics are expected to show up in the effective action
and are crucial for a correct description of the unperturbed quantum vacuum of the fluid. 
However the expectation values $\langle T_{\mu\nu}\rangle$ entering the backreaction equations 
(\ref{eqbackA}, \ref{eqbackB}) 
do not represent the energy momentum of the fluid quantum vacuum.
They describe instead the perturbation of the stationary vacuum (whose energy is strictly 
zero \cite{volovik}) induced by inhomogeneities and by the time dependent formation
of the sonic hole which triggers the phonons emission. In this paper we have assumed that these deviations can be computed 
within the low energy theory. This situation is not unusual. 
Casimir effects are well known examples of vacuum disturbances caused by the presence of boundaries.
It happens that the Casimir energy is often (but not always, 
see G. Volovik in Ref. \cite{libro})
independent on the microscopic physics and can be calculated within the framework of the low
energy theory. This happens because, while low frequency modes are reflected by the 
boundaries, for the high energy
ones the wall is transparent.  They produce a divergent contribution to the vacuum energy which 
is canceled by a proper regularization scheme and does not affect the finite result. We have assumed
that a similar decoupling happens for the acoustic black hole.   
The check of our hypothesis would require an analysis of the quantum system within 
the microscopic theory which takes into account
the time dependent non homogeneous formation of the sonic hole. 
This is for the moment beyond computational capability.
\\More efforts are therefore necessary to fully understand the role played by the high-frequency
modes in the above scenarios; in particular what is needed is a proper description of the way these
modes interact with the underlying medium.
\\To this end the study of black hole analogue models seems much more promising, since there the
underlying physics is at least in principle known. Progress in this area may then be used as
``theoretical laboratory'' for ideas to be exported and tested in quantum gravity. 
\\Given the the present limited understanding, our work is of course not the end story but it should
be regarded simply as a first attempt to attack the backreaction problem in sonic black holes, an
attempt performed using available technical tools developed in quantum field theory in curved
space. 
\\It is obvious that more detailed analysis are needed and will hopefully be performed in the not
too distant future. However being the behaviour we have found for the evolution of a sonic hole
physically quite reasonable, we would expect these features to be confirmed. 

\vspace{1cm}
{\bf Acknowledgements}

We thank R. Emparan, L. Garay, T. Jacobson, M. Maio, R. Parentani and J. Russo 
for useful comments and E. Berti for help with the figure. 

\appendix
\section{Appendix I}
In this appendix we give the relevant expressions needed to evaluate the r.h.s. of the backreaction equations.\\
 
The stress tensor components are
\bea
T^{(2)}_{--}&=&-\frac{\hbar}{12\pi}\frac{(c^2-v^2)}{4c^4}
\left[\left(-\frac{\r''}{2\r}+\frac{3}{4}\frac{{\r'}^2}{\r^2}\right)(c^2-v^2)\right.
+\nonumber\\
&&\left. 
+{v'}^2+vv''+\frac{(vv')^2}{c^2-v^2}\right]+\frac{\hbar \k^2}{48\pi}\ , 
\\ T^{(2)}_{++}&=&T^{(2)}_{--}-\frac{\hbar \k^2}{48\pi} \ . 
\eea
The anomalous trace is given by
\be
T^{(2)}=\frac{\hbar}{24\pi}R^{(2)}\ , 
\ee
where
\be
R=-\frac{1}{\r c}\left[\left(\frac{\r''}{\r}-\frac{{\r'}^2}{\r^2}\right)(c^2-v^2)
-\left(\frac{2vv'\r'}{\r}+2{v'}^2+2vv''\right)\right]\ .\ee 
One needs an expansion of the background velocity field till  the fifth order in $\k z$
\be
\label{A1}
v\simeq -c+c\k z-\frac{1}{6}c \k^2 z^2-\frac{11}{36}c \k^3 z^3
+\frac{77}{1080}c\k^4 z^4+\frac{769}{4320}c\k^5z^5\ . 
\ee
Similarly for the energy density
\be
\label{A2}
\r\simeq \r_o e^{-\half}\left(1+\k z-\frac{1}{6}\k^2z^2-\frac{23}{36}\k^3z^3+
\frac{167}{1080}\k^4z^4+\frac{485}{864}\k^5z^5\right)\ . 
\ee
Inserting the expansions (\ref{A1}) and (\ref{A2}) it is possible to obtain the following expressions, entering in the backreaction equations
\bea
G_2&\simeq&\frac{\k^2c^2A_0^2}{24\pi}\left(\frac{1}{3}+\frac{9}{2}\k z
-\frac{49}{5}\k^2 z^2\right)\ , \\
F_2&\simeq&\frac{|D|A_0\k^3}{48\pi}\left(-\frac{649}{60}-\frac{4909}{360}\k z\right)\ , \\
H(z)&\simeq& \frac{\k^3c^2A_0}{48\pi}\left(\frac{109}{60}
+\frac{19501}{360}\k z\right)\ . 
\eea


\section{Appendix II}

The coefficients entering in the solution of the backreaction equations are given as follows
\bea
a_1&=&\left. \dot{f}_0\right|_{t=0}=\frac{1}{3}\gamma \ , \\
a_2&=&\left. \ddot{f}_0\right|_{t=0}=\frac{1189}{120}\gamma c\k \ , \\
a_3&=&\left. \stackrel{\cdot\cdot\cdot}{f}_0\right|_{t=0}=-\frac{10337}{360}\gamma (c\k)^2\ , \\
b_1&=&\left. \dot{f}_1\right|_{t=0}=\frac{9}{2}\gamma\ , \\
b_2&=&\left. \ddot{f}_1\right|_{t=0}=-\frac{16817}{120}\gamma c\k\ , \\
c_1&=&\left. \dot{f}_2\right|_{t=0}=-\frac{304}{15}\gamma \ , 
\eea
where $\gamma=\frac{\k^2c^2A_0}{24\pi}$.
The first time derivatives are calculated from the boundary conditions
\be
\left. \k^i\frac{\dot{f}_i}{i!}\right|_{t=0}=\left(\frac{G_2}{A}\right)_i
\ee
whereas $\left. \ddot{f}_0\right|_{t=0}$ is calculated from eq. (\ref{zeroorder}) evaluated at $t=0$ and 
 $\left. \ddot{f}_1\right|_{t=0}$  is calculated from eq. (\ref{firstorder}) using the former result for $\ddot{f}_0$, i.e.
\bea
\left. \ddot{f}_0\right|_{t=0}&=&c\left(H_0+2\frac{G_2^{(1)}}{A_0}\right)\ , \\
\left. \ddot{f}_1\right|_{t=0}&=&c\left(\frac{H_1}{\kappa}-H_0-2\frac{G_2^{(1)}}{A_0}
+\frac{4}{\k}\left(\frac{G_2}{A}\right)_2\right)\ . 
\eea
Finally, differentiating eq. (\ref{zeroorder}) and evaluating at $t=0$ we have
\be
\left. \stackrel{\cdot\cdot\cdot}{f}_0\right|_{t=0}=2c^2\k
\left[\frac{H_1}{\kappa}-H_0
+\frac{4}{\k}\left(\frac{G_2}{A}\right)_2\right]\ . 
\ee

\end{document}